\documentclass[twocolumn,letter]{jpsj3}

\usepackage{color}
\usepackage{graphicx}
\usepackage{dcolumn}
\usepackage{bm}

\begin{document}

\title{Spatially Inhomogeneous Superconducting State near $H_{\rm c2}$ in UPd$_2$Al$_3$}
\author{Shunsaku~Kitagawa$^{1,}$\thanks{E-mail address: kitagawa.shunsaku.8u@kyoto-u.ac.jp}, 
Ryoichi~Takaki$^{1}$,
Masahiro~Manago$^{1}$,
Kenji~Ishida$^{1}$,
Noriaki~K.~Sato$^{2}$}
\inst{$^1$Department of Physics, Graduate School of Science, Kyoto University, Kyoto 606-8502, Japan \\
$^2$Department of Physics, Graduate School of Science, Nagoya University, Nagoya 464-8602, Japan
}

\date{\today}

\newcommand{\red}[1]{\textcolor{red}{#1}}

\abst{
We have performed $^{27}$Al-NMR measurements on single-crystalline UPd$_2$Al$_3$ with the field parallel to the $c$ axis to investigate the superconducting (SC) properties near the upper critical field of superconductivity $H_{\rm c2}$.
The broadening of the NMR linewidth below 14~K indicates the appearance of the internal field at the Al site, which originates from the antiferromagnetically ordered moments of U 5$f$ electrons.
In the SC state well below $\mu_0H_{\rm c2}$ = 3.4~T, the broadening of the NMR linewidth due to the SC diamagnetism and a decrease in the Knight shift are observed, which are well-understood by the framework of spin-singlet superconductivity. 
In contrast, the Knight shift does not change below $T_{\rm c}(H)$, and the NMR spectrum is broadened symmetrically in the SC state in the field range of 3~T $< \mu_0 H < \mu_0 H_{\rm c2}$.
The unusual NMR spectrum near $H_{\rm c2}$ suggests that a spatially inhomogeneous SC state such as the Fulde-Ferrell-Larkin-Ovchinnikov (FFLO) state would be realized.
}


\abovecaptionskip=-5pt
\belowcaptionskip=-10pt

\maketitle

Superconductivity in the presence of magnetic fields close to the upper critical field ($H_{\rm c2}$) is the critical state where the intriguing physical phenomena are anticipated.
In general, there are two well-known pair-breaking mechanisms in a type-II superconductor under magnetic fields.
One is the orbital pair-breaking effect related to the emergence of Abrikosov vortices, and the superconductivity is destroyed at the vortex cores.
The other is the Pauli pair-breaking effect, which originates from the Zeeman splitting of spin-singlet Cooper pairs.
When the Zeeman-splitting energy is as large as the condensation energy of superconductivity, the superconductivity becomes unstable and transitions to the normal state.
The strong contribution of the Pauli pair-breaking effect leads to many interesting issues, such as the realization of the spatially modulated superconducting (SC) state, which is called the Fulde-Ferrell-Larkin-Ovchinnikov (FFLO) state\cite{P.Fulde_PR_1964,A.I.Larkin_JETP_1965}.
For the realization of the FFLO state, a very clean system is necessary; thus, not many compounds have been pointed out as candidates for the FFLO state.

\begin{figure}[!t]
\vspace*{10pt}
\begin{center}
\includegraphics[width=8.5cm,clip]{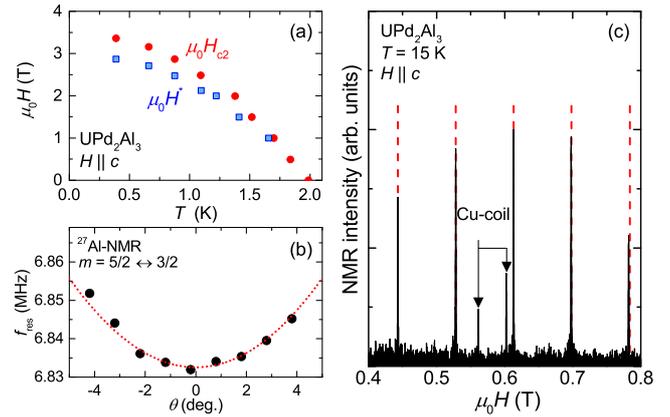}
\end{center}
\caption{(Color online)(a) $H-T$ phase diagram of UPd$_2$Al$_3$ for $H \parallel c$\cite{K.Gloos_PRL_1993}.
Circles represent $\mu_0H_{\rm c2}$, and squares represent $H^*$.
A pronounced hysteresis behavior has been observed in a narrow field range below $H_{\rm c2}$, $H^* < H < H_{\rm c2}$, in several different experiments.
(b) The angular dependence of the resonance frequency of second satellite peak ($m = 5/2 \leftrightarrow 3/2$).
The fitting by a theoretical calculation within first perturbation is indicated by the dotted line.
(c) Field-swept NMR spectrum at 15~K for UPd$_2$Al$_3$ with the field along $c$ axis.
The signal indicated by the solid arrows originates from the Cu coil.
The broken lines are the simulation of the NMR spectrum for $H \parallel c$ using the quadrupole parameters in a previous report\cite{Y.Kohori_SSC_1995}. 
}
\label{Fig.1}
\end{figure}

UPd$_2$Al$_3$ is one such candidate for the FFLO state.
UPd$_2$Al$_3$ has a hexagonal PrNi$_2$Al$_3$-type structure with the space group $P6/mmm$ and exhibits antiferromagnetic (AFM) order at the N\'eel temperature $T_{\rm N}$ = 14.5~K with a commensurate wave vector $Q_{\rm AF} = (0, 0, 0.5)$ and well-localized magnetic moment of $\sim0.85~\mu_{\rm B}$/U\cite{A.Krimmel_ZPB_1992,A.Krimmel_SSC_1993}.
After the AFM transition occurs, superconductivity is observed below the SC transition temperature $T_{\rm c}$ = 2~K\cite{C.Geibel_ZPB_1991}.
NMR and $\mu$SR measurements provide evidence for a spin-singlet Cooper pairing from the clear decrease in the Knight shift below $T_{\rm c}$\cite{M.Kyogaku_PhyiscaB_1993,R.Feyerherm_PhysicaB_1994}.
Large specific-heat jump at $T_{\rm c}$\cite{C.Geibel_ZPB_1991} indicates that heavy electrons form Cooper pairs; thus, the orbital pair-breaking field becomes large.
As a result, the suppression of superconductivity by the Pauli pair-breaking effect is expected.
Moreover, a pronounced hysteresis behavior has been observed in a narrow field range below $H_{\rm c2}$, $H^* < H < H_{\rm c2}$, with several different experiments, such as magnetization, ac-susceptibility\cite{T.Liihmann_PhysicaC_1994,Y.Haga_JPSJ_1996}, ultrasound-velocity\cite{R.Modler_PhysicaB_1993}, and thermal-expansion measurements\cite{K.Gloos_PRL_1993}, as shown in Fig.~\ref{Fig.1} (a).
These experimental results have been considered to be indirect evidence of the existence of the FFLO state in UPd$_2$Al$_3$; however, there is no microscopic evidence of the realization of the FFLO state until now.

In this paper, we report the results of $^{27}$Al-NMR measurements under magnetic field parallel to the $c$ axis, which were performed in order to investigate the SC properties near $H_{\rm c2}$ microscopically.
We found the anomalous NMR spectrum near $H_{\rm c2}$; the Knight shift does not change even below $T_{\rm c}(H)$, but the NMR spectra are broadened symmetrically in the SC state above 3~T.
The spectrum suggests that the spin susceptibility in the SC state near $H_{c2}$ is spatially inhomogeneous, which is consistent with the FFLO state.  

A 47-mg single crystal of UPd$_{2}$Al$_{3}$ was prepared from a carefully homogenized melt of high-purity elements by the Czochralski method in a triarc furnace\cite{N.Sato_JPSJ_1992}.
The single crystal was of very high quality, as determined by $T_{\rm c} = 2$ K and the size of the jump in the specific heat of $\Delta C/T_{\rm c} = 140$ mJ / (mol K$^2$).
A conventional spin-echo technique was utilized for the following NMR measurements.
$^{27}$Al-NMR spectra (a nuclear spin $I$ = 5/2 and a nuclear gyromagnetic ratio $^{27}\gamma/2\pi = 11.094$~MHz/T) were obtained as a function of the frequency in a fixed magnetic field.
The $^{27}$Al nuclear spin?lattice relaxation rate $1/T_1$ was determined by fitting the time dependence of the spin-echo intensity after saturation of the nuclear magnetization to a theoretical function.
Low-temperature NMR measurements down to 100~mK were carried out with a $^3$He-$^4$He dilution refrigerator, in which the single-crystal sample is immersed into the $^3$He-$^4$He mixture to avoid rf heating during the measurements.

In the NMR measurement, the degeneracy of the nuclear spin degrees of freedom is lifted by the Zeeman ($\mathcal{H}_{\rm Z}$) and electric quadrupole ($\mathcal{H}_{\rm Q}$) interactions.
The total effective Hamiltonian is expressed as follows:
\begin{align}
\mathcal{H} &= \mathcal{H}_{\rm Z} + \mathcal{H}_{\rm Q} \notag \\
            &= -\gamma \hslash (1 + K)I \cdot H + \frac{\nu_{zz}}{6}\left\{(3I_z^2-I^2)+\frac{1}{2}\eta(I_+^2+I_-^2)\right\},
\end{align}
where $K$ is the Knight shift, $H$ is the external field, $\nu_{zz}$ is the quadrupole frequency along the principal axis of the electric field gradient (EFG) and is defined as $\nu_{zz} \equiv 3e^2qQ/2I(2I-1)$ with $eq = V_{zz}$,  $eQ$ is the electric quadrupole moment, and $\eta$ is the asymmetry parameter of the EFG expressed as $(V_{xx}-V_{yy})/ V_{zz}$, where $V_{\alpha \alpha}$ is the second derivative of the electric potential $V$ and is the EFG along the $\alpha$ direction ($\alpha = x,y,z$).
It has been reported that $\nu_{zz} = 5.9$~MHz parallel to the $c$ axis and $\eta~=~0.2$\cite{Y.Kohori_SSC_1995}.
Six nuclear spin levels were well-separated; thus, we observed five resonance lines, as shown in  Fig.~\ref{Fig.1} (c).
The widths of the five lines are comparable to each other.
Since the position of a resonance line depends on the angle between the applied magnetic field and the principal axis of the EFG ($c$ axis), we can estimate the field direction with respect to the $c$ axis from the NMR peak locus.
The angle dependence of resonance frequency $f_{\rm res}$ for the second satellite ($m = 5/2 \leftrightarrow 3/2$) peak of the $^{27}$Al-NMR spectrum can be expressed as
\begin{align}
f_{\rm res} = \gamma [1+K(\theta)] H - \nu_{zz}(3\cos^2\theta - 1 - \eta \sin^2\theta \cos 2\phi)
\end{align}
within the first-order perturbation.
Here, $\theta$ and $\phi$ are the elevation angle between the $c$ axis and the applied field and the azimuth angle between the $a$ axis and the $ab$-plane component of the applied field, respectively.
We used a single-axis rotator for the alignment of the  magnetic field direction, and the misalignment of the $c$ axis with respect to the field-rotation plane was estimated to be less than 1$^{\rm o}$ from the analysis of the NMR spectrum, as shown in Fig.~\ref{Fig.1} (b).
A tiny signal from the impurity phase was observed near the central peak.
Then, $^{27}$Al-NMR measurements were performed at the first satellite ($m = 3/2 \leftrightarrow 1/2$) peak above 2~K and at the second satellite ($m = 5/2 \leftrightarrow 3/2$) peak below 2~K to avoid the effect of the impurity signal.
Since the width of the NMR signal is similar for the five lines, the accuracy of the analysis did not change for these lines.

\begin{figure}[!b]
\vspace*{10pt}
\begin{center}
\includegraphics[width=8.5cm,clip]{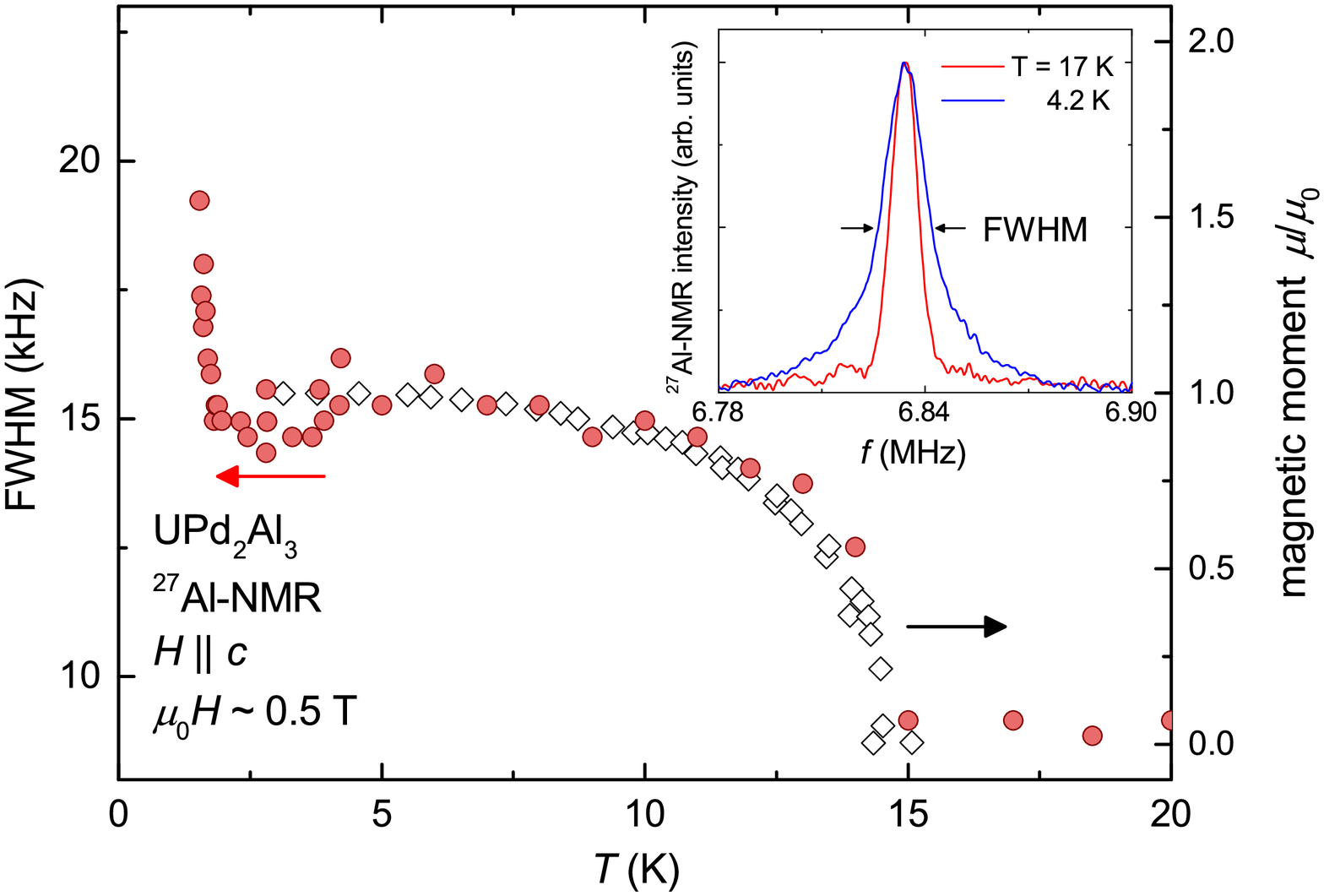}
\end{center}
\caption{(Color online) Temperature dependence of full width at half maximum (FWHM) around 0.5~T together with magnetic moment estimated from neutron scattering\cite{R.Feyerherm_PhysicaB_1994}.
The measurements were performed at the first satellite ($m = 3/2 \leftrightarrow 1/2$) peak.
(inset) Frequency-swept $^{27}$Al-NMR spectra at 17~K and 4.2~K for $H \parallel c \sim 0.5$~T.}
\label{Fig.2}
\end{figure}

\begin{figure*}[!tb]
\vspace*{10pt}
\begin{center}
\includegraphics[width=17cm,clip]{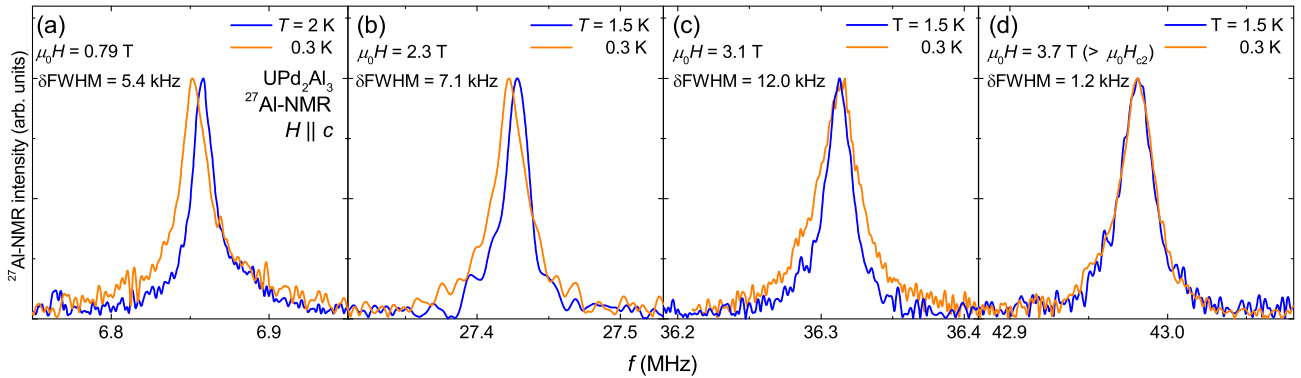}
\end{center}
\caption{(Color online) Frequency-swept $^{27}$Al-NMR spectra above and below $T_{\rm sc}$ at different fields.
The measurements were performed at the second satellite ($m = 5/2 \leftrightarrow 3/2$) peak.
$\delta$FWHM $\equiv$ FWHM (0.3~K) $-$ FWHM ($T \sim T_{\rm c}$) in each spectrum is represented.
At 3.7~T [greater than $\mu_0H_{\rm c2}(0)$], the $^{27}$Al-NMR spectra at 0.3~K and 1.5~K are shown.}
\label{Fig.3}
\end{figure*}
Figure~\ref{Fig.2} shows the temperature dependence of the full width at half maximum (FWHM) of the $^{27}$Al-NMR spectrum measured at $\sim$0.5~T down to 1.5~K.
As shown in the inset of Fig.~\ref{Fig.2}, the linewidth of the $^{27}$Al-NMR spectrum increased below $T_{\rm N}$, indicating the appearance of a small internal field along the $c$ axis at the Al site.
The temperature dependence of the FWHM was scaled to that of the magnetic moment determined from the neutron scattering measurement\cite{R.Feyerherm_PhysicaB_1994}, indicating that the linewidth of the $^{27}$Al-NMR spectrum reflects the AFM moments at the U site.
However, it is noted that the transferred hyperfine field from the ordered U moments should cancel at the magnetically symmetric Al site in the AFM state with $Q_{\rm AF}$ = (0, 0, 0.5). Moreover, the internal field was not observed from the $\mu$SR measurements, where a muon stops at the symmetric (0, 0, 1/2) site between two U ions along the $c$ axis\cite{R.Feyerherm_PhysicaB_1994}. 
Therefore, it is considered that the appearance of the internal field is caused by the imperfect cancellation of the hyperfine fields from the U moments due to the small displacement of the Al atoms from the ideal position. 
The appearance of the internal field at the Al site was also observed in previous $^{27}$Al-NQR measurements\cite{Y.Kohori_SSC_1995}.
Since small displacement from the ideal position was reported in heavy-fermion superconductor UPt$_3$\cite{D.Walko_PRB_2001}, this might be often observed in U-based hexagonal compounds.

\begin{figure}[!b]
\vspace*{10pt}
\begin{center}
\includegraphics[width=7cm,clip]{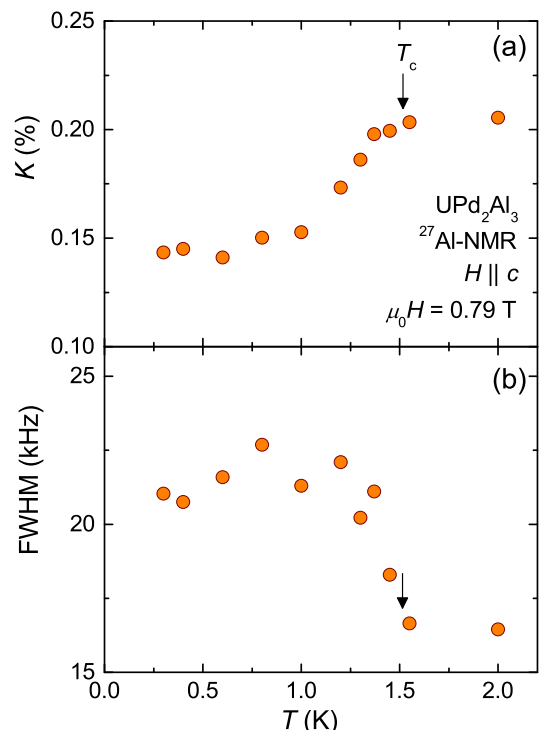}
\end{center}
\caption{(Color online)Temperature dependence of the (a) Knight shift and (b) FWHM at $\mu_0H = 0.79$~T.
The solid arrows indicate $T_{\rm c}$.}
\label{Fig.4}
\end{figure}

\begin{figure}[!tb]
\vspace*{10pt}
\begin{center}
\includegraphics[width=8cm,clip]{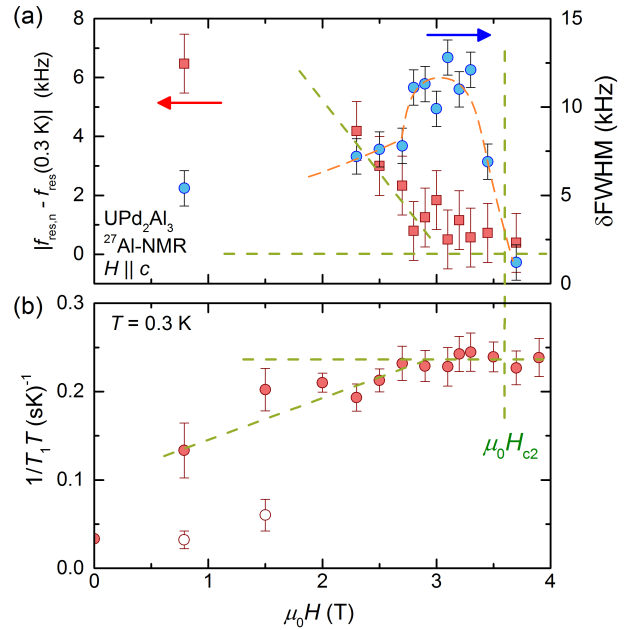}
\end{center}
\caption{(Color online)Magnetic field dependences of (a) the absolute value of the shift in the resonance frequency and the variation in the FWHM, and (b) $1/T_1T$ at 0.3~K (b) for $H \parallel c$.
The broken lines and solid curve are guides for the eyes.}
\label{Fig.5}
\end{figure}

In the SC state well below $H_{\rm c2}$, the broadening of the linewidth and a decrease in the resonance frequency were observed, as shown in Figs.~\ref{Fig.3} (a) and (b).
Figure \ref{Fig.4} shows the temperature dependences of the Knight shift and linewidth of the NMR spectrum shown in Fig.~\ref{Fig.3} (a).
The broadening is due to the SC diamagnetic shielding effect, and the decrease in the resonance frequency indicates the decrease in the Knight shift due to spin-singlet pair formation.
Both phenomena have been observed in a conventional spin-singlet superconductor, and are analyzed later. 
In contrast, the NMR spectra near $H_{\rm c2}$ showed quite unusual behavior.
The resonance frequency did not change even below $T_{\rm c}(H)$, but the NMR spectra were broadened symmetrically, as shown in Fig.~\ref{Fig.3} (c).
As shown in Fig.~\ref{Fig.3} (d), this symmetrical broadening was not observed above $H_{\rm c2}$ at all, indicating that the unusual broadening of the NMR spectra is a characteristic feature of the SC state near $H_{\rm c2}$.
To clarify the magnetic field region in which this anomalous broadening is observed, the magnetic field dependence of the variation in the FWHM, $\delta$FWHM $\equiv$ FWHM (0.3~K) $-$ FWHM ($T \sim T_{\rm c}$), and the absolute value of the resonance frequency shift, $|\delta f|~=~|f_{\rm res}({\rm normal~state}) - f_{\rm res}(0.3~{\rm K})|$, are shown in Fig.~\ref{Fig.5} (a).
$\delta$FWHM increased with increasing $H$ and shows maximum in the field range between 2.8~T and 3.2~T.
In addition, $|\delta f|$ decreased with increasing magnetic field owing to the destruction of superconductivity by the magnetic field, and $|\delta f|$ was almost zero around 3~T, although $\mu_{0}H_{\rm c2}$ is 3.4~T at 0.3~K.
This field range is quite consistent with the $H^*$ anomaly, where pronounced hysteresis behavior was observed in various measurements\cite{T.Liihmann_PhysicaC_1994,Y.Haga_JPSJ_1996,R.Modler_PhysicaB_1993,K.Gloos_PRL_1993}. 

We also measured $1/T_1$ at 0.3 K and found the anomalous behavior in the field region of $H^* < H < H_{c2}$.  
Figure~\ref{Fig.5} (b) shows the magnetic field dependence of $1/T_1$ at 0.3~K.
The value of $1/T_1$ obtained by the NQR measurement was quite small since the quasiparticle excitation was suppressed by the formation of an SC gap, and no vortices were induced at zero field.      
At 0.8 and 1.5~T, the recovery curve of the nuclear magnetization after the saturation pulse can be fitted by a theoretical function with two different $T_1$ components. 
The smaller $1/T_1$ indicated by the open circles represents $1/T_1$ of the SC region far from the vortices, and the larger $1/T_1$ indicated by the closed circles represents $1/T_1$ near the vortex core. 
This is the reason why the larger component of $1/T_1$ is close to the normal-state value. 
However, such two-component behavior became blurred in the field above 1.5~T since the difference between two components became smaller with increasing $H$; thus, the fitting with one $T_1$ component was carried out above 1.5~T.         
As seen in Fig.~\ref{Fig.5} (b), $1/T_1$ increased with increasing $H$, which was interpreted as the increase in the field-induced normal state, and $1/T_1$ reached the same value as $1/T_1$ in the normal state already around $H^*$, and remained almost constant up to 4~T. 
These results also suggest the presence of the anomalous SC state in $H^* < H < H_{\rm c2}$. 

Now, we discuss the SC properties of UPd$_2$Al$_3$ quantitatively.
First, we discuss whether or not the decrease in the Knight shift in the SC state is a reasonable value with respect to the Pauli-limiting field $H_{\rm P}(0)$, since only one-fourth of the total susceptibility decreased in the SC state.
It is well-known that, in a spin-singlet superconductor, a relation holds between the Pauli-limiting field and the change in the spin susceptibility $\delta \chi$ ascribed to singlet-pair formation, which is expressed as,
\begin{align*}
\frac{1}{2}\delta\chi \mu_0H_{\rm P}(0)^2~=~\frac{1}{2}(\chi_{\rm N} -\chi_{\rm SC})\mu_0H_{\rm P}(0)^2~=~\frac{1}{2}\mu_0H_{\rm c}^2,
\end{align*}
where $\chi_{\rm N}$ and $\chi_{\rm SC}$ represent the spin susceptibilities in the normal and the SC states, respectively and $H_{\rm c}$ is the critical field of superconductivity. Thus, $\delta~\chi~=~\chi_{\rm N} -\chi_{\rm SC} $.
This equation leads to $\mu_0H_{\rm P}(0) = \mu_0H_{\rm c}/\sqrt{\chi_{\rm N}-\chi_{\rm SC}}$.
$\chi_{\rm N}-\chi_{\rm SC}$ can be determined from the difference between the Knight shift in the normal state and that in the SC state.
As shown in Fig.~\ref{Fig.4} (a), the difference in the Knight shifts for the SC and normal states is approximately $\delta K = 0.08~\%$; thus, $\chi_{\rm N}-\chi_{\rm SC}$ along the $c$ axis is estimated to be 0.13 $\times$ 10$^{-2}$ emu/mol from the relation of $\delta \chi = (\mu_B N_A / ^{27}A_{\rm hf}) \delta K$ with the hyperfine coupling constant at the Al site $^{\rm 27}A_{\rm hf}$ =  3.6~kOe/$\mu_B$, which was estimated from the linear relation between $\chi$ and $K$\cite{Y.Kohori_SSC_1995}. 
By using the same relation to the decrease in the Knight shift perpendicular to the $c$ axis, which was measured with a field-aligned polycrystalline sample\cite{M.Kyogaku_PhyiscaB_1993}, $\chi_{\rm N}-\chi_{\rm SC}$ perpendicular to the $c$ axis is estimated to be 0.19 $\times$ 10$^{-2}$ emu/mol.   
From the critical field $H_{\rm c}$ = 814~Oe of UPd$_2$Al$_3$\cite{Y.J.Uemura_PhysicaB_1993}, $\mu_0H_{\rm P}(0)$ is estimated to be 5.1~T for $H \parallel c$ and 4.2~T for $H \perp c$, which is smaller than the orbital-limiting field $\mu_0H_{\rm c2}^{\rm orb}$ = 6.0~T and not so far from the observed upper critical fields $\mu_0H_{\rm c2}$ = 3.5~T for $H \parallel c$ and 3.0~T for $H \perp c$.
These results indicate that the dominance of the Pauli effect near $H_{c2}$ is also shown from the decrease in the spin susceptibility measured by the Knight shift. 
The predominance of the Pauli effect near $H_{c2}$ is one of the necessary condition for the realization of the FFLO state. 

Next, we consider the origin of the broadening of the NMR spectrum below $T_{\rm c}(H)$.
In general, the broadening of the NMR linewidth due to the vortex effect in type-II superconductors is expressed as,
\begin{align}
\delta {\rm FWHM(H)} = \frac{\Phi_0}{4\sqrt{\pi^3}\lambda_{ab}^2}\left(1-\frac{H}{H_{c2}^{\rm orb}}\right),
\end{align}
where $\lambda_{ab}$ is the penetration depth in the $ab$ plane.
Using $\lambda_{ab}~=~4600$~\AA\cite{R.Feyerherm_PhysicaB_1994} and $\mu_0H_{\rm c2}^{\rm orb}$ = 6.0~T, the field distribution in the vortex state is estimated to be $\delta$FWHM(0.5~T) = 0.36~mT $\sim$ 4.1~kHz, which is consistent with the observed $\delta$FWHM at low fields ($\delta$FWHM $\sim 5$~kHz).
This conventional vortex effect should decrease with increasing $H$, but $\delta$FWHM is larger near $H_{\rm c2}$.
It is obvious that $\delta$FWHM near $H_{c2}$ cannot be explained by this effect.
In addition, the NMR spectrum is broadened on both the lower- and higher-frequency sides, and the spectrum becomes symmetric above 3~T, as shown in Fig.~\ref{Fig.3} (c).
Since the resonance frequency reflects the internal fields at the nuclear site, the broadening on both the lower- and higher-frequency sides implies that the magnetic susceptibility $\chi$ of some part of the sample is larger or smaller than that in the normal state, and these parts equally exist in the sample.
It was revealed from our NMR experiments that the SC state above 3~T becomes an inhomogeneous SC state.  

Here, we consider the origin of such an inhomogeneous SC state above 3~T. 
There are two possibilities that induce the symmetrical broadening of the NMR spectrum below $T_{\rm c}(H)$.
One is that the observed distribution of the internal field is ascribed to the increase in the U-AFM moments in the SC state.
As shown in Fig.~\ref{Fig.2}, the increase in $\delta$FWHM of the $^{27}$Al NMR spectrum originates from the evolution of the U-AFM moments, and if the U-AFM moments increase in the SC state, $\delta$FWHM increases below $T_{\rm c}(H)$.
However, it is estimated that the AFM moments should increase by nearly double in the SC state from the linear relation between  $\delta$FWHM and the U-AFM moments in Fig.~\ref{Fig.2}; thus, this possibility would be excluded since an increase in the AFM moments has not been reported\cite{H.Kita_JPSJ_1994}.
The other possibility is that the distribution of the internal field originates from the SC properties.
$\chi$ in the SC region becomes smaller owing to the spin-singlet pairing, but $\chi$ in the vortex region becomes larger than $\chi$ in the normal state, probably owing to a strong paramagnetic effect.
The experimentally observed symmetrical line broadening means that the large and small $\chi$ regions have nearly the same volume, suggesting the presence of a large vortex region.
A large vortex region is predicted in the FFLO state.
However, it is noted that a strong enhancement in 1/$T_1T$ just below $H_{\rm c2} (T)$, which is regarded as a hallmark of the FFLO state in $\kappa-$(BEDT-TTF)$_{2}$Cu(NCS)$_{2}$\cite{H.Mayaffre_NatPhys_2014}, was not observed for UPd$_{2}$Al$_{3}$.
It was shown from a theoretical study\cite{B.M.Rosemeyer_PRB_2016} that the enhancement in $1/T_1T$ originates from the presence of the Andreev bound state inside the SC gap, and that the magnitude of the enhancement depends on the SC gap structure.
Therefore, we consider that the different behavior of $1/T_1T$ could be interpreted with the difference in SC gap structures of $\kappa-$(BEDT-TTF)$_{2}$Cu(NCS)$_{2}$ and UPd$_{2}$Al$_{3}$.
To investigate the unusual SC state near $H_{\rm c2} (T)$, $\bm{k}$-resolved measurements such as neutron scattering experiments or scanning tunneling spectroscopy measurements are needed to get a direct evidence of the FFLO state since NMR is a local probe and cannot provide $\bm{k}$-dependent information.

In conclusion, we performed $^{27}$Al-NMR measurements on single-crystalline UPd$_2$Al$_3$ with the magnetic field applied along the $c$ axis to investigate the SC properties near $H_{\rm c2}$.
The broadening of the NMR linewidth below 14~K indicates the appearance of the internal fields originating from the antiferromagnetically ordered moments.
In the SC state well below $\mu_0H_{\rm c2}$, an additional broadening of the NMR linewidth and a decrease in the resonance frequency are observed, which can be understood by the framework of conventional spin-singlet superconductivity.
On the other hand, the resonance frequency does not change even below $T_{\rm c}(H)$, and the NMR spectra are broadened symmetrically in the SC state above 3~T.
This suggests that a spatially inhomogeneous SC state is realized near $H_{\rm c2}$, determined by the Pauli-depairing effect.
We insist on the reconsideration of the possibility of the FFLO state realized in UPd$_{2}$Al$_{3}$.

\section*{Acknowledgments}
The authors acknowledge S. Yonezawa, Y. Maeno, and Y. Matsuda for fruitful discussions. 
This work was partially supported by the Kyoto Univ. LTM Center and Grant-in-Aids for Scientific Research (KAKENHI) (Grant Numbers JP15H05882, JP15H05884, JP15K21732, JP15H05745, and JP17K14339). 


\end{document}